\documentclass[prp,twocolumn,superscriptaddress,floatfix,showpacs]{revtex4}
\usepackage{graphicx,amsfonts,amssymb,amsmath, hyperref,dsfont}

\newif\ifhyper
\hypertrue
\ifhyper
\hypersetup{
   citecolor = {green},
   colorlinks = {true}, 
   urlcolor = {blue} 
}
\fi

\newcommand{\beq}{\begin{equation}}
\newcommand{\eeq}{\end{equation}}
\newcommand{\beqa}{\begin{eqnarray}}
\newcommand{\eeqa}{\end{eqnarray}}

\def\Longarrow{\protect\@lra}
\def\@lra{\relbar\joinrel\relbar\joinrel\relbar\joinrel%
          \relbar\joinrel\rightarrow}

\begin{document}
\title{Quantum disorder and local modes of the fully-frustrated transverse field Ising model on a diamond chain}

\author{K.~Coester}
\affiliation{Lehrstuhl f\"ur Theoretische Physik 1, TU Dortmund, Germany}
\author{W.~Malitz}
\affiliation{Lehrstuhl f\"ur Theoretische Physik 1, TU Dortmund, Germany}
\author{S.~Fey}
\affiliation{Lehrstuhl f\"ur Theoretische Physik 1, TU Dortmund, Germany}
\author{K.~P.~Schmidt}
\email{kai.schmidt@tu-dortmund.de}
\affiliation{Lehrstuhl f\"ur Theoretische Physik 1, TU Dortmund, Germany}

\date{\rm\today}

\begin{abstract}
We investigate the transverse field Ising model on a diamond chain using series expansions about the high-field limit
 and exact diagonalizations. For the unfrustrated case we accurately determine the quantum critical point and its expected 2d Ising universality separating the polarized and the $\mathcal{Z}_2$ symmetry broken phase. In contrast, we find strong evidences for a disorder by disorder scenario for the fully-frustrated transverse field Ising model, i.e.~except for the pure Ising model, having an extensive number of ground states, the system is always in a quantum disordered polarized phase. The low-energy excitations in this polarized phase are understood in terms of exact local modes of the model. Furthermore, an effective low-energy description for an infinitesimal transverse field allows to pinpoint the quantum disordered nature of the ground state via mapping to an effective transverse field Ising chain and to determine the induced gap to the elementary effective domain wall excitation very accurately.  
\end{abstract}


\maketitle

\section{Introduction}\label{sec:introduction}

Geometrically frustrated quantum magnets display a fascinating variety of quantum ground states and associated emergent phenomena. Typically, frustration gives rise to an extensive number of ground states in the classical limit and the addition of quantum fluctuations leads to exotic collective quantum behavior. Perhaps the simplest and hence most tractable models featuring such an interplay of geometric frustration and quantum fluctuations are the fully-frustrated Ising systems in a transverse field \cite{liebmann1986}. On lattices like the square or the honeycomb lattice, the frustration results from the incompatibility of the odd number of ferro- and anti-ferromagnetic bonds in a single plaquette having an even number of bonds. 

All fully-frustrated transverse field Ising models (TFIMs) exhibit a classical ground-state manifold with a macroscopic degeneracy which is lifted in the presence of an infinitesimal transverse field inducing quantum fluctuations. In the limit of high transverse fields, all TFIMs realize a quantum disordered polarized phase. The situation for the opposite limit of weak fields is much more involved. A large number of frustrated TFIMs exhibits a nontrivial quantum ordered phase arising from classical disorder, a scenario which is also known as \emph{order by disorder} \cite{kano1953,villain1980,shender1982}. In contrast, the kagome lattice plays a distinguished role in two dimensions as it is very reluctant to order \cite{moessner2000,moessner2001}. This system is a quantum paramagnet for arbitrary transverse fields providing an instance of \emph{disorder by disorder} \cite{powalski13} in which quantum fluctuations select a quantum disordered state out of the classically disordered ground-state manifold. The same kind of disorder by disorder scenario has recently been established for the topologically ordered toric code in a field on the dice lattice using a duality transformation \cite{schmidt13}. 

To date, the only known TFIM featuring such an unusual disorder by disorder scenario in one dimension is the sawtooth chain \cite{priour2001}, a system of cornersharing triangles. The central objective of this paper is to give strong evidences for a disorder by disorder scenario for the frustrated TFIM on a diamond chain. The latter system consists of cornersharing squared plaquettes with an even number of bonds and it therefore represents the first instance of disorder by disorder for such a system. 

This paper is organized as follows. In Sec.~\ref{sec:model} we briefly introduce the TFIM on a diamond chain. In Sec.~\ref{sec:methods} we give technical details for the high-order series expansions and the exact diagonalizations. The unfrustrated and frustrated TFIM are studied in Sec.~\ref{sec:UFTFIM} and in Sec.~\ref{sec:FFTFIM}. An effective low-energy description in the limit of infinitesimal fields is presented in Sec.~\ref{sec:low_field}. Finally, Sec.~\ref{sec:summary} provides a concluding summary.

\section{Model}
\label{sec:model} 
We study the Ising model in a transverse magnetic field on a diamond chain. The corresponding Hamiltonian is given by
\begin{equation}
\label{eq:ising}
	H= \sum_{<i,j>}J_{ij}^{\phantom{z}}\sigma^z_i \sigma^z_j-h^{\phantom{x}}\sum_i{\sigma^x_i}
\end{equation}
with Pauli matrices $\sigma_i^{\alpha}$ acting on site $i$, the antiferromagnetic nearest-neighbor exchange $|J_{ij}|=J$, and the strength of the transverse magnetic field $h$. Here we consider the two cases $\prod_{\langle i,j\rangle\in p} J_{ij}/J =\pm 1$ on every plaquette $p$, where $+1$ is the unfrustrated TFIM and $-1$ is the fully-frustrated TFIM on the diamond chain. 

\begin{figure}[htbp]
	\centering
		\includegraphics[width=1.0\columnwidth]{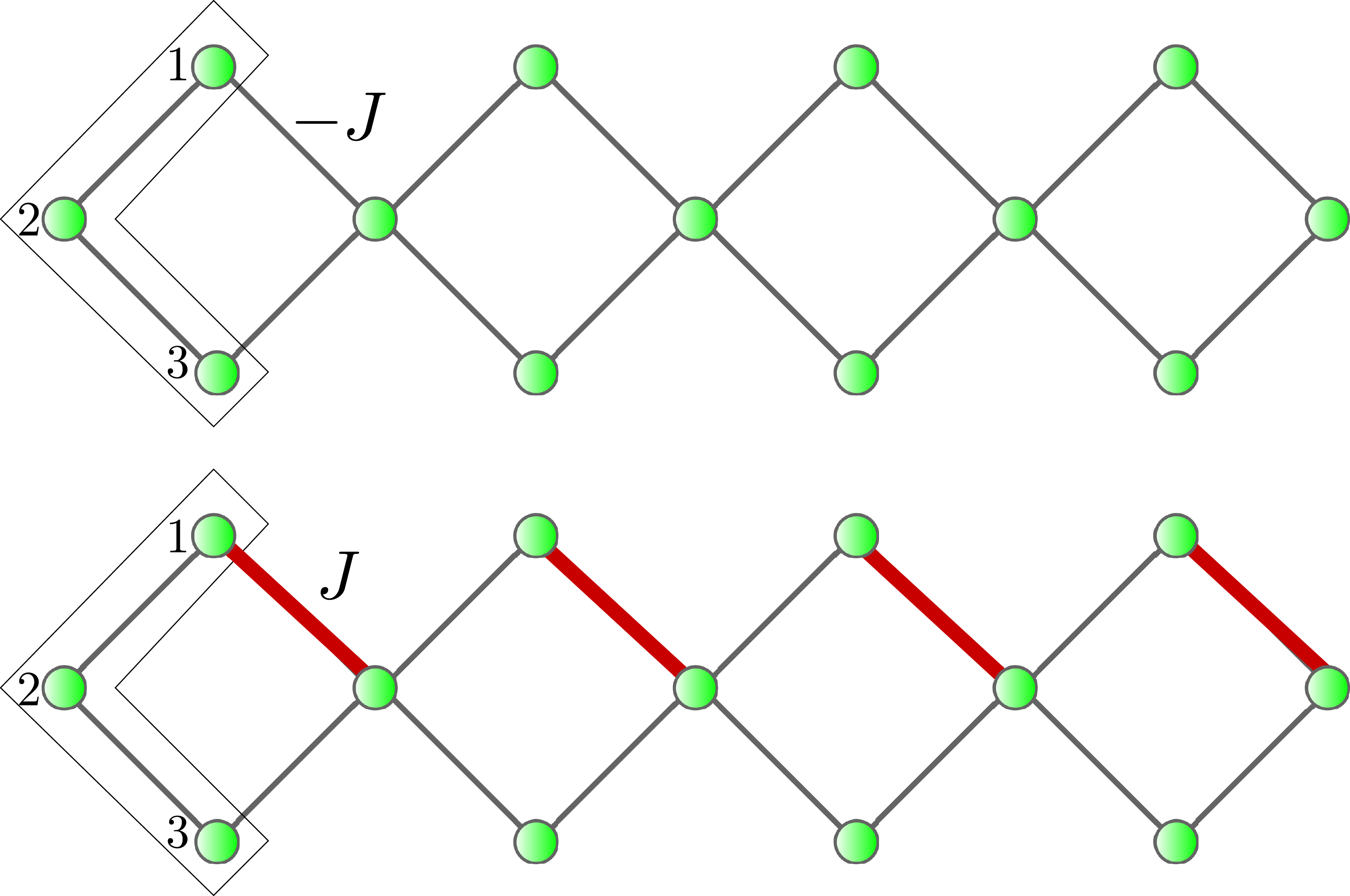}
	\caption{Illustration of the TFIM on a diamond chain. If all Ising couplings $J_{\rm ij}=-J$ are equal, the system is unfrustrated (upper panel). The system is highly frustrated if the Ising coupling on the thick red bonds for each plaquette $p$ has a different sign compared to the other three bonds in the same plaquette (lower panel). In both cases one can use the three-site unit cell with sites 1, 2, and 3 illustrated as a black box.}
	\label{Fig:diamond_chain}
\end{figure}

To be more specific, we consider the following two choices of Ising couplings as illustrated in Fig.~\ref{Fig:diamond_chain}
\begin{eqnarray}
 H_{\rm uf/ff} &=& -J \sum_{\nu} \Big( \sigma^z_{\nu,2} \sigma^z_{\nu,1}+\sigma^z_{\nu,2} \sigma^z_{\nu,3}+\sigma^z_{\nu,3} \sigma^z_{\nu+1,2} \nonumber\\
               &\pm& \sigma^z_{\nu,1} \sigma^z_{\nu+1,2} \Big)-h\sum_i{\sigma^x_i}\quad ,
\end{eqnarray} 
where $H_{\rm uf}$ ($H_{\rm ff}$) corresponds to the unfrustrated (frustrated) case and $\nu$ denotes the unit cell consisting of three sites $1$, $2$, and $3$. In the following we focus on $J\geq 0$.

\section{Method} 
\label{sec:methods}
Starting from the high-field limit $h\gg J$ , we use perturbative continuous unitary transformations (pCUTs) \cite{knetter2000,knetter2003} to derive a quasi-particle conserving effective Hamiltonian in the zero- and one-particle sector up to high order in perturbation. 

For this purpose, it is expedient to interpret the elementary excitations corresponding to single spin flips as quasi-particles above the vacuum which is given by the fully polarized state. The spin flips are then described in terms of hardcore bosons represented by creation and annihilation operators $b_i^{\dagger}$ and $b_i^{\phantom{\dagger}}$. Besides the usual bosonic commutation relations these operators meet the hardcore constraint
\begin{equation}
	  b_i^{\phantom{\dagger}} b_i^{\phantom{\dagger}} = b_i^{\dagger} b_i^{\dagger} = 0 \quad 
\end{equation}
which means that every site $i$ can be occupied at most by one boson as every spin is either 'up' or 'down' (or in bosonic language: 'occupied' or 'empty').
 
The Hamiltonian (\ref{eq:ising}) can then be written as
\begin{eqnarray}
	\frac{H}{2h} &=& E_0 + \sum_i \hat{n}_i + \sum_{\langle i,j \rangle }x_{ij}\left( b^\dagger_i b^{\phantom{\dagger}}_j + b^\dagger_i b^\dagger_j + {\rm h.c.}\right)   \nonumber\\
                    &=&H_0 + x \left( T_0+T_{+2}+T_{-2} \right)= H_0 + x V \label{eq:initial_hamiltonian}
\end{eqnarray}
where $|x_{ij}|=x=J/2h$ is the perturbation parameter, \mbox{$E_0=-N/2$}, and $\hat{n}_i = b^\dagger_i b^{\phantom{\dagger}}_i$ is the local density on site $i$. Here $N$ is the total number of sites. We identify the Ising term as the perturbation $V$ which can be decomposed into a sum of $T_m$ operators incrementing (or decrementing) the number of quasi-particles by $m=\left\{+2,-2,0 \right\}$. 

The pCUT maps the initial Hamiltonian (\ref{eq:initial_hamiltonian}) to an effective Hamiltonian of the form
\begin{equation}
H_\text{eff}=H_0+\sum_{k=1}^{\infty} x^k \sum_{|\underline{m}|=0} C(\underline{m}) T_{m_1} \dots T_{m_k} \label{eq:eff_hamiltonian}
\end{equation}
where $\underline{m}=(m_1, \dots, m_k)$ and the coefficients $C(\underline{m})$ are rational numbers. The constraint \mbox{$|\underline{m}|=m_1+\dots + m_k=0$} reflects the quasi-particle conserving property of the effective Hamiltonian. The terms $T_{m_1} \dots T_{m_k}$ can be interpreted as virtual fluctuations involving up to $k$ bonds. Due to the linked cluster theorem only linked fluctuations have a finite overall contribution. Hence, the effective Hamiltonian (\ref{eq:eff_hamiltonian}) can be used to calculate the one-particle excitation energies in the thermodynamic limit.

The unfrustrated and the frustrated diamond chain have a three-site unit cell. The effective one-particle Hamiltonian can therefore be written as
\begin{eqnarray}
 H^{\rm 1qp}_{\rm eff} &=& \sum_k \sum_{\alpha,\beta} \omega^{\alpha\beta}(k) b^\dagger_{k,\alpha} b^{\phantom{\dagger}}_{k,\beta} \\
                       &=& \sum_k \sum_{\alpha} \bar{\omega}^{\alpha}(k) b^\dagger_{k,\alpha} b^{\phantom{\dagger}}_{k,\alpha} \quad .     
\end{eqnarray}
In this definition we have introduced the annihilation and creation operators in momentum space defined by \mbox{$b^{(\dagger )}_{k,\alpha}=(1/\sqrt{N_{\rm uc}})\sum_\nu\,e^{\pm{\rm i}k\nu}\,b^{(\dagger)}_{\nu,\alpha}$} and $\alpha,\beta\in\{1,2,3\}$ specify the site inside the unit cell. Note, that the number of unit cells $N_{\rm uc}$ is equal to $N/3$.

One therefore has three one-particle bands $\bar{\omega}^{\alpha}(k)$ for both diamond chains which we have calculated as high-order series expansions in the perturbative parameter $x$. The minimal one-particle energy $\Delta\equiv {\rm min}(\bar{\omega}^{\alpha}(k),\{ k,\alpha\})$ corresponds to the one-particle gap which is the central quantity we want to analyze in the following sections.

\section{Unfrustrated TFIM}
\label{sec:UFTFIM}

In this section we study the zero-temperature phase diagram of the unfrustrated TFIM $H_{\rm uf}.$ For $h=0$, one has two degenerate ground states, either all spins point up or down. In the opposite limit $J=0$, the ground state is unique corresponding to the fully polarized state where all spins point in $x$ direction. In between these two limits, one expects a second-order phase transition in the 2D Ising universality class separating the polarized high-field phase from the $\mathcal{Z}_2$ symmetry broken phase present at small fields.

We have calculated the one-particle bands $\bar{\omega}^{\alpha}(k)$ as high-order series expansions up to order 12 in $x$. Let us stress that due to the reflection symmetry of the unfrustrated diamond chain with respect to the centerline of the chain (see Fig.~\ref{Fig:diamond_chain}), the $3\times 3$ matrix $\omega^{\alpha\beta}(k)$ is block-diagonal having an isolated band $\bar{\omega}^{1}_{\rm uf}(k)$ with odd parity and a $2\times 2$ subblock with even parity.

\begin{figure}[htbp]
	\centering
		\includegraphics[width=1.0\columnwidth]{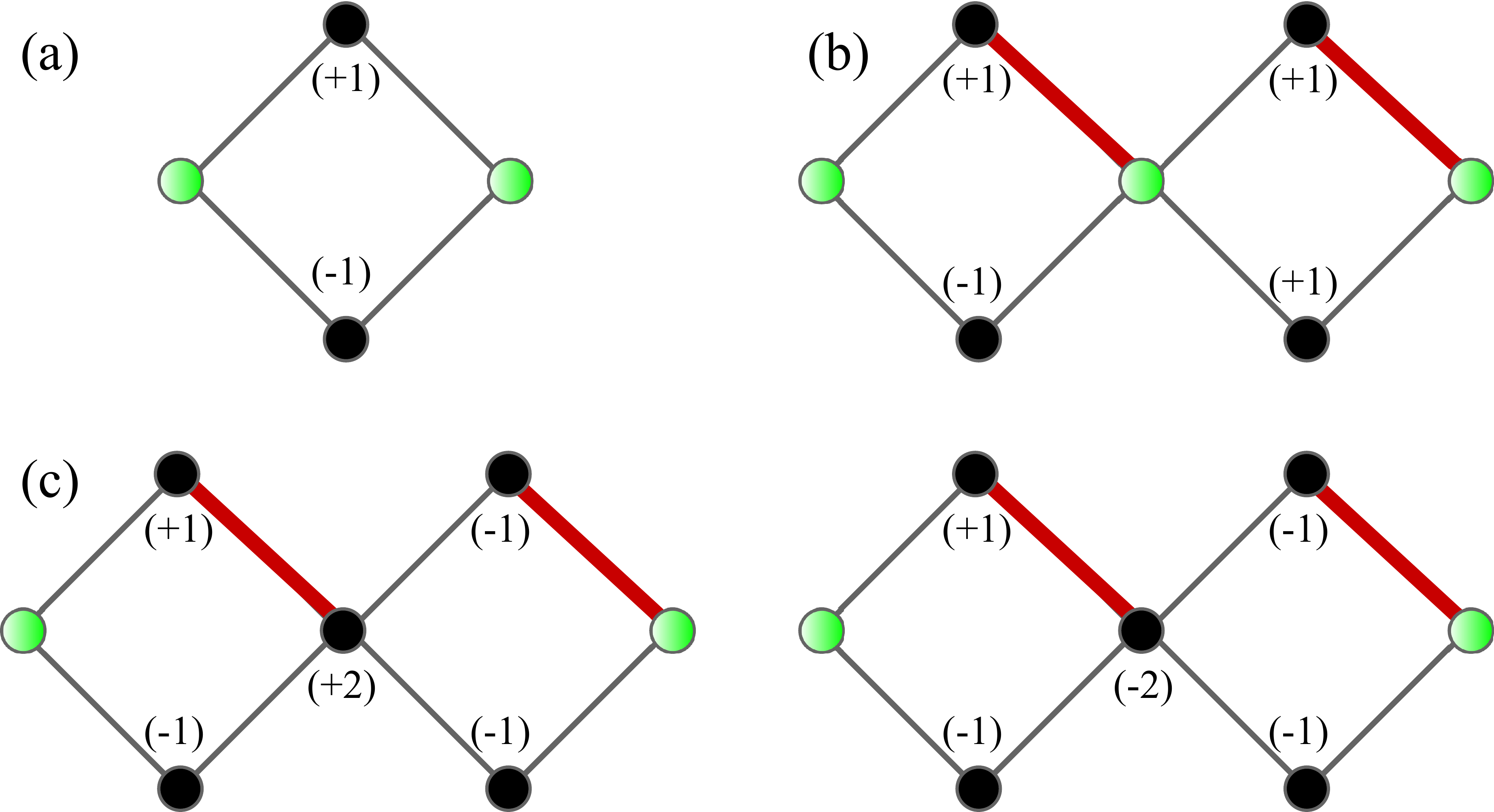}
	\caption{Illustration of the local modes of the (a) unfrustrated and (b-c) frustrated diamond chain.}
	\label{Fig:local_modes}
\end{figure}

The band $\bar{\omega}^{1}_{\rm uf}(k)$ is found to be exactly flat which can be understood in terms of a local one-particle mode (see also Ref.~\onlinecite{schulenburg2002} for the physics of localized magnons) 
\begin{equation}
 |\nu\rangle_1^{\rm uf} = \frac{1}{\sqrt{2}}\left( |\nu,1\rangle - |\nu,3\rangle\right)
\end{equation}
with $H_{\rm eff}|\nu\rangle_1^{\rm uf}=\bar{\omega}_1|\nu\rangle_1^{\rm uf}$. Here, $|\nu,\alpha\rangle$ denotes a state with one particle on site $\alpha$ of unit cell $\nu$. This mode is illustrated in Fig.~\ref{Fig:local_modes}a. The locality results from exact destructive interferences due to the opposite amplitudes $\pm 1$ in the definition of $|\nu\rangle_1^{\rm uf}$ following exactly the same principles already established in Ref.~\onlinecite{powalski13}. However, contrary to the results on the kagome lattice, there are no fluctuations breaking the destructive interference pattern and the mode is local up to infinite order perturbation theory.

The bands with even parity have a finite dispersion starting in leading order perturbation theory. The one-particle gap $\Delta_{\rm uf}$ is located at $k=0$ and it has even parity. Up to order 12, the gap is given by
\begin{eqnarray}
 \Delta_{\rm uf}&=& 1-2\sqrt{2}x-x^2-\frac{9\sqrt{2}}{8}x^3+\frac{3}{2}x^4-\frac{15\sqrt{2}}{256}x^5\nonumber\\
              &&-\frac{61}{8}x^6+\frac{63367\sqrt{2}}{4096}x^7-\frac{227}{4}x^8\nonumber\\
              &&+\frac{84785637\sqrt{2}}{262144}x^9-\frac{443355}{512}x^{10}\nonumber\\
              &&+\frac{25959361825\sqrt{2}}{4194304}x^{11}-\frac{31389385}{2048}x^{12}\, .
\end{eqnarray}

\begin{figure}[htbp]
	\centering
		\includegraphics[width=1.0\columnwidth]{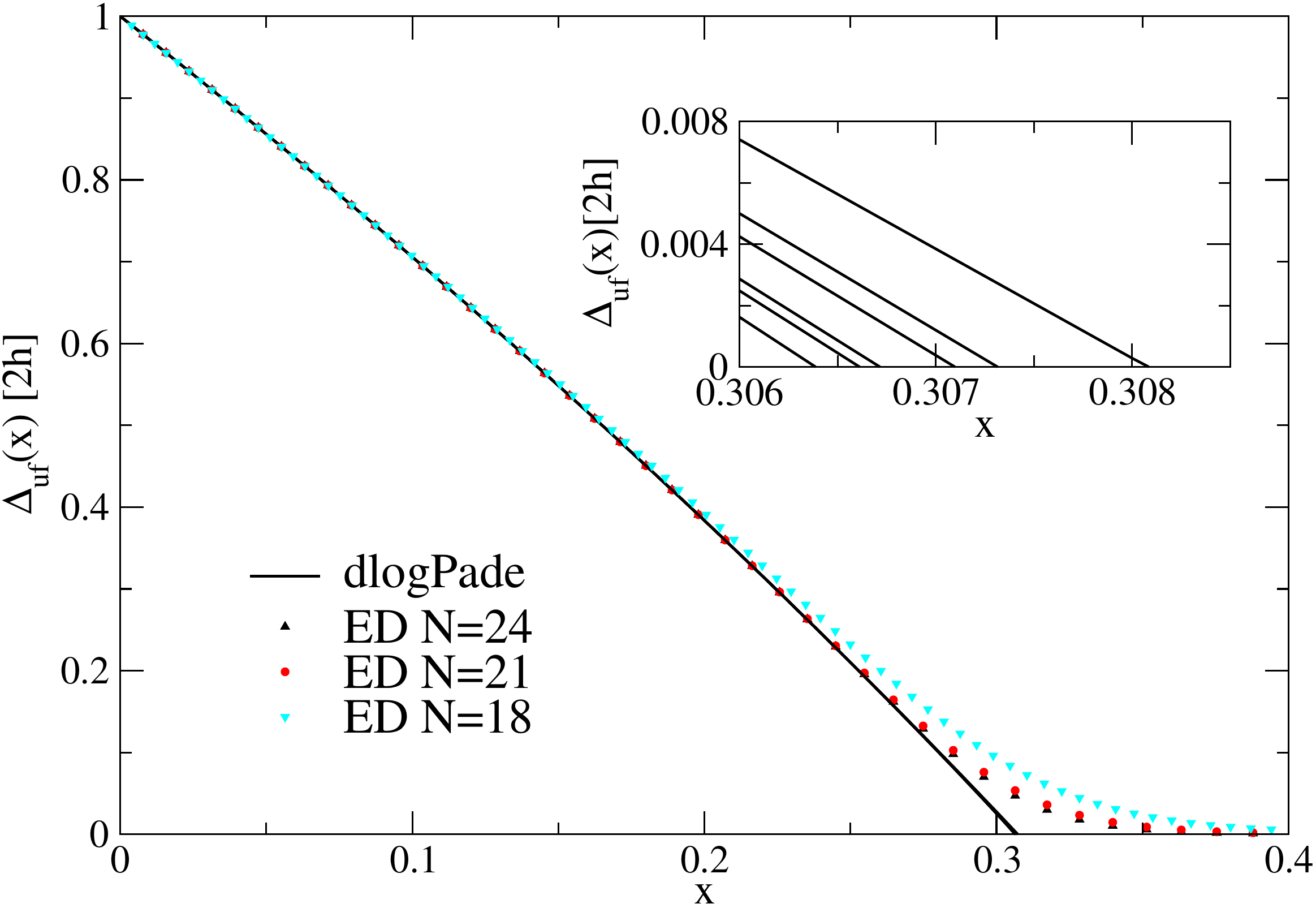}
	\caption{One-particle gap $\Delta_{\rm uf}/2h$ as a function of $x$ for the unfrustrated TFIM. Lines correspond to dlogPad\'e extrapolants while symbols represent exact diagonalization data of different finite-size samples with $N$ sites. {\it Inset}: Zoom on the critical region.}
	\label{Fig:UFTFIM_x_extrapolation}
\end{figure}

Apart from the one-particle gap $\Delta_{\rm uf}$ obtained by series expansions, we use exact diagonalizations of
 finite systems to calculate the gap by taking the difference between the first two energy levels of finite
 systems up to $N=24$ sites. All results are shown in Fig.~\ref{Fig:UFTFIM_x_extrapolation}. 
  
We find that many DlogPad\'{e} approximants are defective displaying an unphysical pole at intermediate values of $x$. Restricting to the remaining approximants yields convincingly a quantum critical point at $J/2h\approx 0.307(1)$. This is also in good agreement with the exact diagonalization data. Furthremore, the critical exponent $z\nu$ is found to be $\approx 0.99$ from the DlogPad\'{e} approximants fully consistent with the expected 2D Ising universality class having $z=\nu=1$. 

\section{Frustrated TFIM}
\label{sec:FFTFIM}

In this section we aim at analyzing the fate of the polarized phase for the frustrated diamond chain. This is again done by comparing high-order series expansions of the one-particle excitations and exact diagonalizations. 

We find that all three one-particle bands $\bar{\omega}_\alpha (k)$ are exactly flat for the frustrated diamond chain. The latter is found for the series expansion in the thermodynamic limit as well as for the lowest excitation energies calculated by exact diagonalization on finite clusters. This exact momentum independence can be understood (again) by means of the high-field limit in terms of local modes which are illustrated in Fig.~\ref{Fig:local_modes}(b-c). For this purpose we define
\begin{eqnarray}
 |\bar{\nu}\rangle_1 &=& \frac{1}{2}\Big( |\nu,1\rangle -|\nu,3\rangle + |\nu+1,1\rangle +|\nu+1,3\rangle\Big) \nonumber\\
 |\nu\rangle_{2/3} &=& \frac{1}{\sqrt{8}}\Big( |\nu,1\rangle -|\nu,3\rangle - |\nu+1,1\rangle \nonumber\\
                   && -|\nu+1,3\rangle \pm 2 |\nu +1,2\rangle \Big) \quad .
\end{eqnarray} 
The resulting modes are local up to infinite order perturbation theory and the guiding priniciples are similar to those in Ref.~\onlinecite{powalski13}. Yet, the sign of the couplings $\pm J$ play an essential role in the resulting locality of these modes and must be taken into account. This is exemplified in first-order perturbation theory for the local mode $|\bar{\nu}\rangle_1$: $\left(H_0+x T_0\right)|\bar{\nu}\rangle_1 =|\bar{\nu}\rangle_1$. The destructive interference leftwards is caused by the different signs of the wave functions whereas the destructive interference rightwards is caused by the different signs of the couplings. This precept can be extended to infinite order perturbation theory, i.e.~the mode $|\bar{\nu}\rangle_1$ is an exact local mode such that $H_{\rm eff}|\bar{\nu}\rangle_1=\bar{\omega}_1|\bar{\nu}\rangle_1$.

This  is different for $|\nu\rangle_{2/3}$.  The two other local modes $|\bar{\nu}\rangle_{2,3}$ are linear combinations of $|\nu\rangle_2$ and $|\nu\rangle_3$. Interestingly, we find that \mbox{$_2\langle\nu|H_{\rm eff}(x)|\nu\rangle_{2}$=$_3\langle\nu|H_{\rm eff}(-x)|\nu\rangle_{3}$}. In contrast, the off-diagonal components \mbox{$_2\langle\nu|H_{\rm eff}|\nu\rangle_{3}$=$_3\langle\nu|H_{\rm eff}|\nu\rangle_{2}$} have only entries of even order. As a consequence, the relation $\bar{\omega}_{2}(x)=\bar{\omega}_{3}(-x)$ holds.

Up to order 14, the energies of the three flat bands $\bar{\omega}_\alpha$ are given by
\begin{eqnarray}
	 \bar{\omega}_1 &=& 1-14x^2+4x^4+\frac{5}{4}x^6-\frac{145}{8}x^8+\frac{116665}{512}x^{10}\nonumber\\
                        &&-\frac{4861423}{6144}x^{12}-\frac{367688725}{442368}x^{14} \\
	 \bar{\omega}_{2,3} &=& 1\mp2x+x^2\pm\frac{11}{4}x^3-\frac{7}{2}x^4\pm\frac{217}{64}x^5-\frac{53}{4}x^6\nonumber\\
                        &&\pm\frac{3315}{512}x^7+\frac{1499}{64}x^8\pm\frac{425909}{16384}x^9+\frac{18211}{1024}x^{10}\nonumber\\
                        &&\mp\frac{70854467}{131072}x^{11}+\frac{6035581}{6144}x^{12}\nonumber\\
                        &&\mp\frac{32007715867}{18874368}x^{13}+\frac{1614585673}{442368}x^{14} \quad .
\end{eqnarray}

In the following we analyze the one-particle gap \mbox{$\Delta_{\rm ff}\equiv \bar{\omega}_2$} on the whole parameter axis in order to clarify whether the frustrated TFIM on a diamond chain realizes a disorder by disorder scenario. At this point we assume that binding effects are small for the TFIM and, consequently, multi-particle sectors play no role for the excitation gap (for any $J/h$). This assumption is reasonable since the leading contribution to the two-particle interaction is repulsive and of order $n=2$ perturbation theory. Additionally, this is confirmed below by exact diagonalizations. 

\begin{figure}[htbp]
	\centering
		\includegraphics[width=1.0\columnwidth]{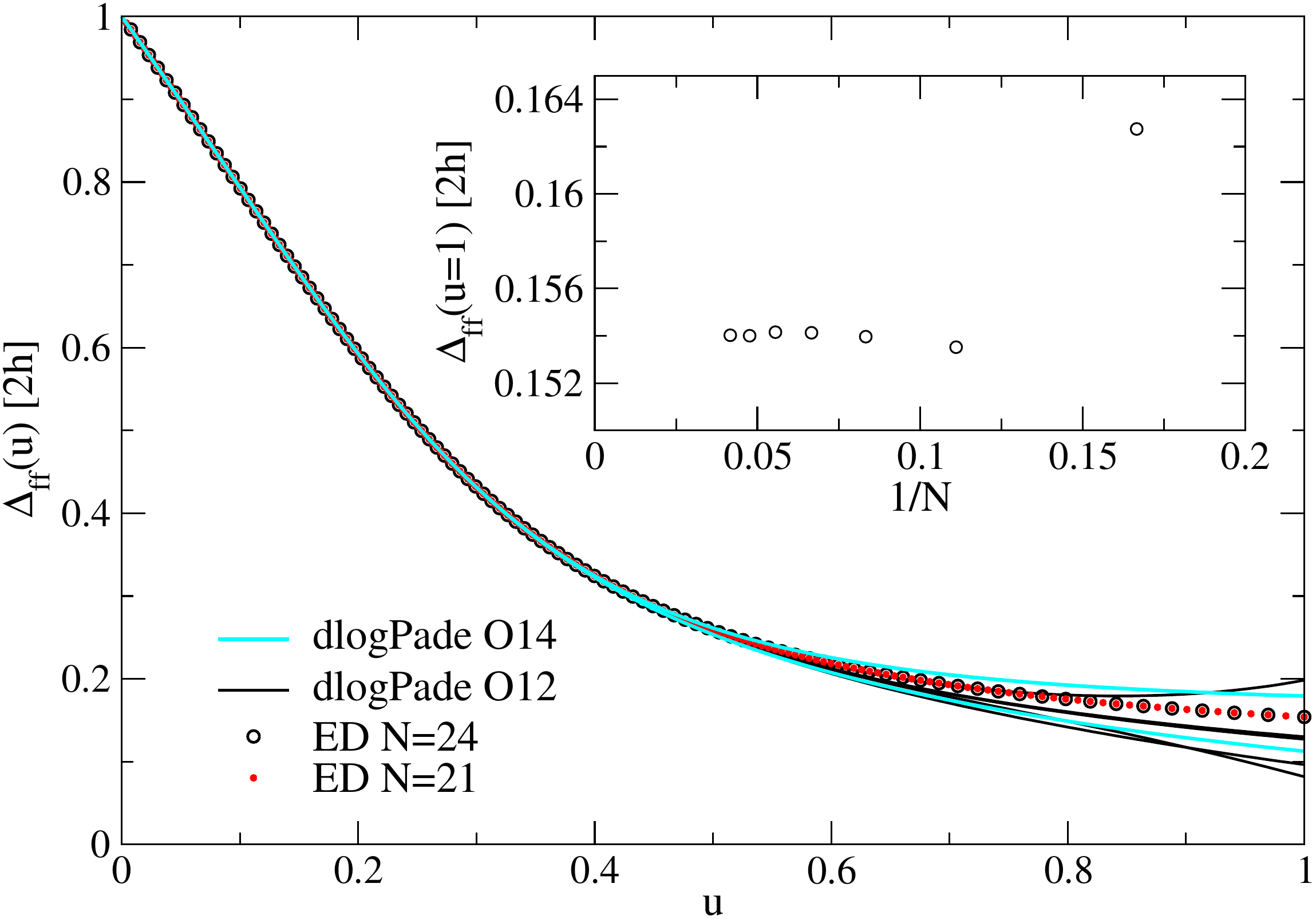}
	\caption{One-particle gap $\Delta_{\rm ff}/2h$ as a function of $u$. Lines correspond to dlogPad\'e extrapolants in $u=x/(x+1)$ while symbols represent exact diagonalization data for different finite-size samples. {\it Inset}: Limiting value $\Delta(u=1)/2h$ from ED as a function of $1/N$.}
	\label{Fig:FFTFIM_u_extrapolation}
\end{figure}

For the extrapolation it is convenient to perform an Euler transformation 
\begin{equation}
	x =: \frac{u}{1-u}
\end{equation}
which maps the parameter axis $x\in {[0,\infty[} $ on the finite interval $u\in [0,1]$. One expects that the dlogPad\'e extrapolants in $u$ are more stable within the finite interval $u\in{[0,1]}$ allowing to estimate the gap for $x\rightarrow \infty$ which corresponds to $u=1$. The resulting extrapolations of the high-field gap are shown in Fig.~\ref{Fig:FFTFIM_u_extrapolation}.

For $u<0.6$, all extrapolations are well converged. A closing of the gap in this region can therefore be excluded. Note that the bare series of $\Delta_{\rm ff}$ converges very slowly as the absolute value of the coefficients grows exponentially with ascending order and, additionally, as the sign of the coefficients alternate. This behavior is very similar to the analogue series for the kagome TFIM \cite{powalski13}. But also far beyond the high-field regime most extrapolants indicate the clear trend that there is no tendency to close the gap for any value of $u$. Additionally, most of the extrapolations approach a finite value $\Delta_{\rm ff}\approx 0.15 \pm 0.05$ for $u=1$. 

These findings are fully confirmed by our ED results which are also displayed in Fig.~\ref{Fig:FFTFIM_u_extrapolation}. Moreover, it turns out that the elementary particle gap is already fully converged in ED on the full parameter axis for finite systems up to $N=24$ sites. Indeed, even for $u=1$ one finds a value $0.154(1)$ with almost no finite-size effects (see inset of Fig.~\ref{Fig:FFTFIM_u_extrapolation}).   

One can therefore conclude without any doubt that the frustrated TFIM on the diamond chain realizes a disorder by disorder 
scenario at zero temperature. In comparison to the other two cases, namely the sawtooth chain and the kagome lattice, the numerical accuracy
 is very high for the diamond chain. This is likely a consequence of the fact that the all one-particle excitations correspond to exact local modes for the diamond chain which is not the case for the other two systems.

\section{Low-field limit}\label{sec:low_field}
The perfect convergence of the frustrated TFIM on the diamond chain, signaling a disorder by disorder scenario, motivates an investigation of the low-field limit $h\ll |J|$ in order to gain a deeper physical understanding. As shown above, an infinitesimal field lifts the extensive ground-state degeneracy of the pure Ising model such that the ground state is unique and the elementary excitation is still $N_{\rm uc}$-fold degenerate which is associated with the local mode $|\bar{\nu}\rangle_2$. The gap in the low-field limit $\tilde{\Delta}_{\rm ff}$ is connected to the gap of the high-field limit by
\begin{equation}
\frac{\partial}{\partial h} \tilde{\Delta}_{\rm ff}(h)|_{h=0}= \lim_{J\rightarrow \infty} \Delta_{\rm ff}(J)|_{h=1}\quad .
\end{equation}
In the following, we use first-order degenerate perturbation theory about the low-field limit to pinpoint the quantum disordered nature of the ground state and to calculate the gap $\lim_{J\rightarrow \infty} \Delta_\text{ff}(J)|_{h=1}$ accurately.

For the unperturbed problem $h=0$, one has the extensive number $4^{N_{\rm uc}}$ of ground states. In fact, any state is a ground state if it has exactly one frustrated bond per plaquette. Here a frustrated bond corresponds to a ferromagnetic (antiferromagnetic) alignment of spins for an antiferromagnetic (ferromagnetic) Ising coupling on a bond.
 
\begin{figure}[htbp]
	\centering
		\includegraphics[width=1.0\columnwidth]{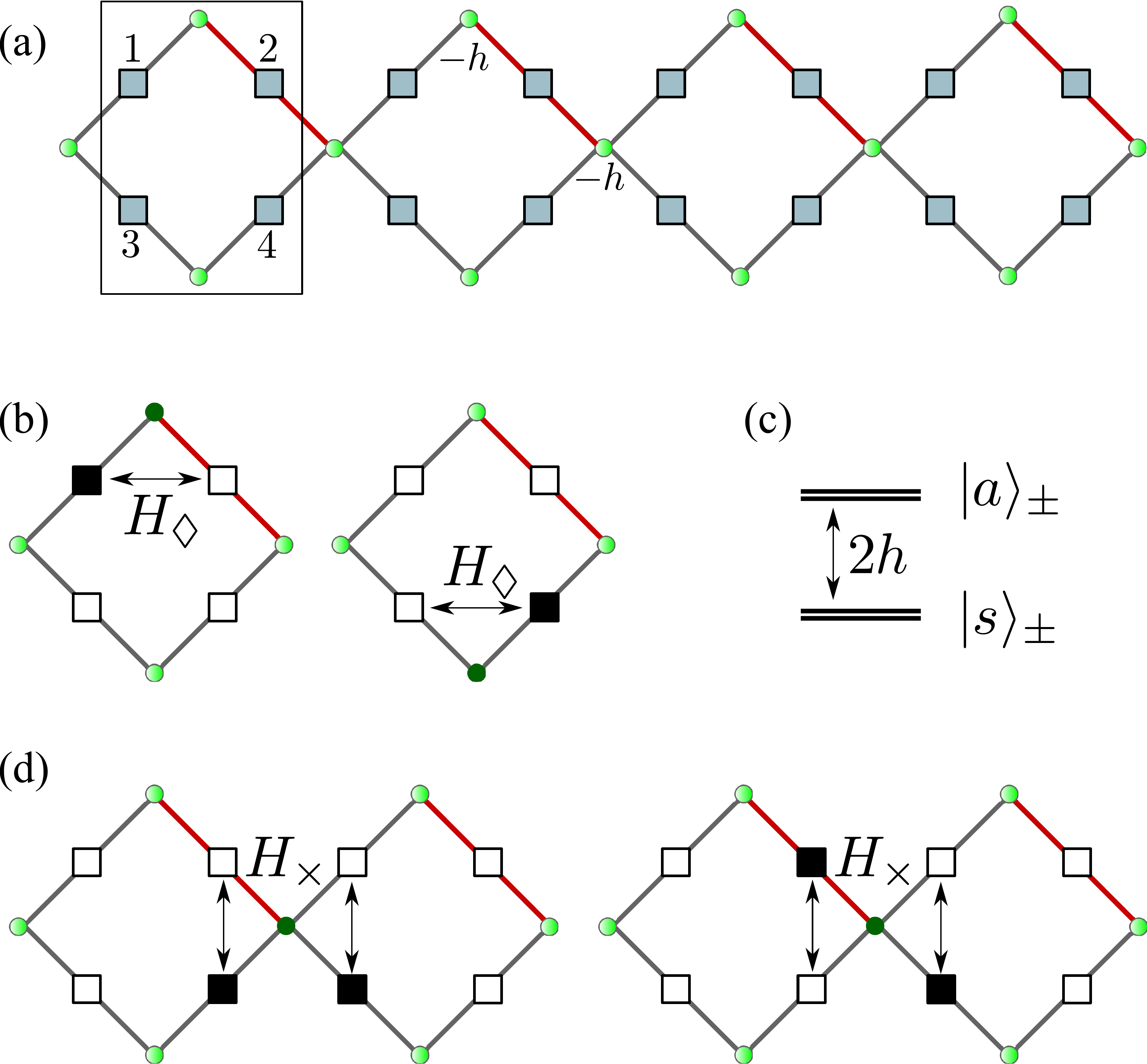}
	\caption{Illustration of the low-energy description in the low-field limit: (a) Frustrated bonds live on the squares which build the effective dual lattice of bonds. The black rectangular represents the unit cell containing the four squares denoted by 1, 2, 3, and 4. (b) Processes of the Hamiltonian $H_\lozenge$ which take place inside a unit cell. Black squares (white squares) correspond to the presence (absence) of a frustrated bond. (c) Local spectrum of $H_\lozenge$ with eigenstates $|s\rangle_\pm$ and $|a\rangle_\pm$. (d) Processes of the Hamiltonian $H_\times$ acting on nearest-neighbor unit cells such that the number of frustrated bonds equals one in each unit cell. }
	\label{Fig:dual_lattice}
\end{figure}

It is then useful to introduce an intuitive picture and to describe the physics in terms of frustrated bonds, i.e.~a frustrated bond is viewed as a quasi-particle which is a hardcore boson living on the dual lattice of bonds as illustrated in Fig.~\ref{Fig:dual_lattice}(a). All states containing exactly one particle (one frustrated bond) per plaquette are ground states of the unperturbed Ising model. Note that ground states related by a global spin flip are mapped onto the same state in the language of frustrated bonds.

In what follows we are deriving an effective Hamiltonian using degenerate perturbation theory in $h/J$ without considering the unimportant constant energy of the unperturbed Ising model. In first order, the effective Hamiltonian contains all processes where different ground states are coupled by the transverse magnetic field $h$. The action of the field on a site of the original model is such that the occupation of particles is inverted on all bonds containing this site. Keeping in mind that the total number of particles in each plaquette must be one, the effective Hamiltonian in terms of frustrated bonds (fb) is given by $H_{\rm fb}=-h\left( H_{\lozenge}+ H_{\times}\right)$ with 
\begin{eqnarray}
H_{\lozenge}&=&\sum_{\tilde{\nu}} \left( \tilde{b}_{\tilde{\nu},1}^\dagger \tilde{b}_{\tilde{\nu},2}^{\phantom{\dagger}} + \tilde{b}_{\tilde{\nu},3}^\dagger \tilde{b}_{\tilde{\nu},4}^{\phantom{\dagger}} + {\rm h.c.}\right) \\ \label{eq:T_0_plaquette_term}
H_{\times}&=&\sum_{\tilde{\nu}} \Big( \tilde{b}_{\tilde{\nu},2}^\dagger\tilde{b}_{\tilde{\nu}+1,1}^\dagger  \tilde{b}_{\tilde{\nu},4}^{\phantom{\dagger}} \tilde{b}_{\tilde{\nu}+1,3}^{\phantom{\dagger}} \nonumber \\ 
&+& \tilde{b}_{\tilde{\nu},2}^\dagger \tilde{b}_{\tilde{\nu}+1,3}^\dagger \tilde{b}_{\tilde{\nu},4}^{\phantom{\dagger}}\tilde{b}_{\tilde{\nu}+1,1}^{\phantom{\dagger}} + {\rm h.c.}  \Big) \quad .\label{eq:T_0_pert_term}
\end{eqnarray}
Here, $\tilde{b}_{\tilde{\nu},\tilde{\alpha}}^\dagger$ $(\tilde{b}_{\tilde{\nu},\tilde{\alpha}}^{\phantom{\dagger}} )$ creates (annihilates) a frustrated bond \mbox{$\tilde{\alpha}\in\{1,2,3,4\}$} on the plaquette $\tilde{\nu}$ (see Fig.~\ref{Fig:dual_lattice}(a)). One has two different contributions $H_{\lozenge}$ and $H_{\times}$ which reflects the fact that one has two different types of sites in the diamond chain. The upper and lower sites $1$ and $3$ are part of two bonds ($H_{\lozenge}$) while the middle sites $2$ belong to four bonds ($H_{\times}$). The different processes are illustrated in Fig.~\ref{Fig:dual_lattice}(b) and Fig.~\ref{Fig:dual_lattice}(d). 

Interestingly, the parity $(\pm)$ in {\it each} plaquette with respect to reflections about the centerline of the diamond chain is conserved for the full Hamiltonian $H_{\rm fb}$. As a consequence, the Hilbert space decouples into $2^{N_{\rm uc}}$ blocks with fixed parities. It is therefore convenient to switch to the diagonal basis of the quadratic part $H_{\lozenge}$ in order to exploit these conservation laws. A single plaquette $\tilde{\nu}$ has four one-particle states yielding the eigenstates 
\begin{eqnarray}
|s\rangle_\pm &:=& \tfrac{1}{2} \left( (|1\rangle +|2\rangle) \pm (|3\rangle + |4\rangle ) \right) \\
|a\rangle_\pm &:=& \tfrac{1}{2} \left( (|1\rangle -|2\rangle) \pm (|3\rangle - |4\rangle ) \right) 
\end{eqnarray}
with the two-fold degenerate eigenenergies 
\begin{eqnarray}
-hH_{\lozenge}|s\rangle_\pm &=& - h|s\rangle_\pm\\
-hH_{\lozenge}|a\rangle_\pm &=& + h|a\rangle_\pm
\end{eqnarray}
as illustrated in Fig.~\ref{Fig:dual_lattice}(c). 

Hence, the ground state of $H_{\lozenge}$ alone is $2^{N_{\rm uc}}$ degenerate which means that the extensive ground-state degeneracy $4^{N_{\rm uc}}$ of the Ising model is only partly lifted. Next, we consider the term $H_\times$ in the diagonal basis of $H_{\lozenge}$. Due to the parity conservation in each plaquette, matrix elements of $H_\times$ acting on nearest-neighbor plaquettes split into four decoupled sectors $++$, $--$, $-+$, and $+-$. For these four sectors one finds the following relations  
\begin{eqnarray}
&&_+\langle j_1|_+\langle j_2|\,H_{\times}\, | i_1\rangle_+ | i_2\rangle_+ \\
&=&+\left(_-\langle j_1|_-\langle j_2|\,H_{\times}\, | i_1\rangle_- | i_2\rangle_- \right)\\
&=&-\left(_-\langle j_1|_+\langle j_2|\,H_{\times}\, | i_1\rangle_- | i_2\rangle_+ \right)\\
&=&-\left(_+\langle j_1|_-\langle j_2|\,H_{\times}\, | i_1\rangle_+ | i_2\rangle_-\right) \quad ,
\end{eqnarray}
where $j_1,j_2,i_1, i_2 \in \{ a,s  \}$ and all other matrix elements vanish because of parity conservation. Thus, the total spectrum of $H_{\rm fb}$ is invariant under the exchange of $+$ and $-$ in each plaquette. 

In each parity sector, one is left with a pseudo-spin $1/2$ $\tau^\alpha$ on each plaquette. One can therefore associate the state $|a\rangle$ with spin up and the state $|s\rangle$ with spin down. Finally, for each parity sector one obtains the following Hamiltonian
\begin{eqnarray}
 \frac{H_{\tau}}{h} &=& \sum_{\tilde{\nu}} \left(\tau_{\tilde{\nu}}^z + \bar{J}_{\tilde{\nu},\tilde{\nu}+1} \left( \mathds{1} + \tau_{\tilde{\nu}}^+ \tau_{\tilde{\nu}+1}^- + \tau_{\tilde{\nu}}^+ \tau_{\tilde{\nu}+1}^+ + {\rm h.c.}\right)\right)\nonumber\\
                   && +\sum_{\tilde{\nu}} \frac{3\bar{J}_{\tilde{\nu},\tilde{\nu}+1}}{2} \left( -\tau_{\tilde{\nu}}^+ + \tau_{\tilde{\nu}+1}^+ +{\rm h.c.}\right)\nonumber\\
                   && +\sum_{\tilde{\nu}} \frac{\bar{J}_{\tilde{\nu},\tilde{\nu}+1}}{2} \left( -\tau_{\tilde{\nu}}^z\tau_{\tilde{\nu}+1}^+ + \tau_{\tilde{\nu}+1}^z\tau_{\tilde{\nu}}^+ +{\rm h.c.}\right) 
\end{eqnarray}
with $\bar{J}_{\tilde{\nu},\tilde{\nu}+1}=-\frac{1}{4}$ if both plaquettes $\tilde{\nu}$ and $\tilde{\nu}+1$ have the same parity ($++$ or $--$) and $\bar{J}_{\tilde{\nu},\tilde{\nu}+1}=\frac{1}{4}$ if parities are different ($+-$ or $-+$). 

Thus, nearest-neighbor plaquettes having the same parity are energetically favored due to the term $\bar{J}_{\tilde{\nu},\tilde{\nu}+1} \mathds{1}$. It is therefore reasonable to assume that the ground state of $H_{\rm fb}$ is in the sector where all parities are either $+$ or $-$ resulting in a two-fold degeneracy. Most importantly, the Hamiltonian $H_{\tau}$ reduces to the conventional transverse field Ising chain in these two parity sectors
\begin{equation}
  -\frac{N_{\rm uc}}{4}+\sum_{\tilde{\nu}} \left(\tau_{\tilde{\nu}}^z -\frac{1}{4}\left( \tau_{\tilde{\nu}}^+ \tau_{\tilde{\nu}+1}^- + \tau_{\tilde{\nu}}^+ \tau_{\tilde{\nu}+1}^+ + {\rm h.c.}\right)\right)\quad ,
\end{equation}
because all operators in $H_{\tau}$ which are asymmetric with respect to interchanging $\tilde{\nu}$ and $\tilde{\nu}+1$ cancel exactly. The transverse field Ising chain can be solved exactly by fermionization \cite{pfeuty70}. For the above prefactors, one finds a disordered (polarized) ground state with an excitation gap $3h/2$ fully consistent with our expectation of a disorder by disorder scenario for the diamond chain. 

Next, we aim at calculating $\tilde{\Delta}_{\rm ff}$ from $H_{\tau}$ which must be located in a different parity sector. Physically, it is reasonable that this sector is the one with finite but minimal number of nearest-neighbor plaquettes having opposite parities. This suggests to consider the parity sector "$\ldots +++---\ldots$" with one "domain wall". The expected low-energy gap $\approx 0.154(1)$ in units of $2h$ calculated in Sec.~\ref{sec:FFTFIM} for the limit $J\rightarrow\infty$ should then correspond to the difference of the ground-state energies of this sector with the one without domain walls.

In order to check this reasoning, we have introduced a parameter $\lambda$ such that $H_{\rm fb}^\lambda=-h\left( H_{\lozenge}+ \lambda H_{\times}\right)$ and we have set up a high-order series expansion in $\lambda$ for this gap. Up to order 16, one finds the following expression

\begin{widetext}
\begin{eqnarray}
 \frac{\tilde{\Delta}_{\rm ff}}{h} &=& \frac{1}{2}\lambda-\frac{1}{4}\lambda^2+\frac{1}{16}\lambda^3+\frac{1}{256}\lambda^4-\frac{13}{1024}\lambda^5+\frac{809}{196608}\lambda^6+\frac{197}{131072}\lambda^7-\frac{203201}{113246208}\lambda^8+\frac{471275}{1358954496}\lambda^9\nonumber\\
                               &&+\frac{46813429}{115964116992}\lambda^{10}-\frac{622323211}{2087354105856}\lambda^{11}+\frac{74289937}{22265110462464}\lambda^{12}+\frac{13474678373}{133590662774784}\lambda^{13}\nonumber\\
                                &&-\frac{339231960130723}{6925339958244802560}\lambda^{14}-\frac{1946153694553}{153896443516551168}\lambda^{15}+\frac{264599575767192599}{11080543933191684096000}\lambda^{16}\quad .
\end{eqnarray}
\end{widetext}
The amplitudes of this series decrease strongly with increasing order which originates from the facts that (i) the gap corresponds to a difference of ground-state energies of two protected parity sectors and (ii) the energy gaps to excited states inside each parity sector are large. As a consequence, we find that the gap $\tilde{\Delta}_{\rm ff}/2h=0.15402(1)$ at $\lambda=1$, which corresponds to original problem $H_{\rm fb}$, is converged in the first four digits. This is in full agreement with the numerical findings of the frustrated TFIM on the diamond chain presented in the previous section. In particular, the domain wall excitation is localized as expected from the existence of the local mode $|\bar{\nu}\rangle_2$ discussed in Sec.~\ref{sec:FFTFIM}.

Moreover, this perfect agreement essentially proves that the true ground state corresponds exactly to the one of the transvserse field Ising chain in the effective low-energy description. As a consequence, we have demonstrated exactly that the frustrated TFIM on the diamond chain realizes a disorder by disorder scenario. 

\section{Summary}\label{sec:summary}
We have studied the unfrustrated and frustrated TFIM on a diamond chain at zero temperature. The unfrustrated model is found to be in the expected 2D Ising universality class separating a $\mathcal{Z}_2$ broken state and a quantum disordered polarized phase. In contrast, convincing evidences have been found that the frustrated diamond chain realizes a disorder by disorder scenario, i.e.~the ground state at infinitesimal fields is adiabatically connected to the high-field polarized phase. Most strikingly, an effective low-energy description in terms of frustrated bonds allows to clarify exactly that the ground state of the frustrated TFIM on the diamond chain is quantum disordered for an infinitesimal field. The elementary excitation in this effective language corresponds to a localized domain wall which is gapped.

The fully-frustrated TFIM on the diamond chain is therefore the third system displaying disorder by disorder. The other two cases are on lattices of cornersharing triangles, namely the one-dimensional sawtooth chain \cite{priour2001} and the two-dimensional kagome lattice \cite{powalski13}. In contrast, the frustrated diamond chain is made of cornersharing squared plaquettes and it is remarkable that all one-particle bands are exactly flat. In our opinion it is therefore important to study further families of frustrated TFIMs on lattices made of cornersharing plaquettes which we left for future studies.   

\section{Acknowledgements}
KPS acknowledges ESF and EuroHorcs for funding through his EURYI.

\end{document}